\documentclass[11pt]{article}

\usepackage{graphicx}

\usepackage[T1]{fontenc}
\usepackage{fleqn}
\usepackage{wrapfig}
\usepackage{amsmath}

\usepackage{a4,multicol}
\usepackage[figuresright]{rotating}

\begin{document}
\begin{normalsize}

\begin{center}

{\bf Probing the Light Speed Anisotropy with respect to the Cosmic
Microwave Background Radiation Dipole }

\end{center}

\vspace{0.2in}

\noindent V.G.Gurzadyan$^{1,2}$, J.-P.Bocquet$^{3}$,
A.Kashin$^{1}$, A.Margarian$^1$, O.~Bartalini~$^4$,
V.~Bellini~$^5$, M.~Castoldi~$^6$, A.~D'Angelo~$^4$,
J.-P.~Didelez~$^7$, R.~Di~Salvo~$^4$, A.~Fantini~$^4$,
G.~Gervino~$^8$, F.~Ghio~$^9$, B.~Girolami~$^9$, A.~Giusa~$^5$,
M.~Guidal~$^7$, E.~Hourany~$^7$, S.~Knyazyan~$^1$,
V.~Kouznetsov~$^{10}$, R.~Kunne~$^7$, A.~Lapik~$^{10}$,
P.~Levi~Sandri~$^{11}$, A.~Lleres~$^3$, S.~Mehrabyan~$^1$,
D.~Moricciani~$^4$, V.~Nedorezov~$^{10}$, C.~Perrin~$^{3}$,
D.~Rebreyend~$^3$, G.~Russo~$^5$, N.~Rudnev~$^{10}$,
C.~Schaerf~$^4$, M.-L.~Sperduto~$^5$, M.-C.~Sutera~$^5$,
A.~Turinge~$^{12}$
%}

\vspace{0.2in}

\noindent $^1$ Yerevan Physics Institute, 375036 Yerevan, Armenia\\
$^2$ ICRA, Dipartimento di Fisica, Universit\`a ``La Sapienza", 00185 Roma, Italy\\
$^3$ IN2P3, Laboratory for Subatomic Physics and Cosmology, 38026 Grenoble, France\\
$^4$ INFN sezione di Roma II and Universit\`a ``Tor Vergata", 00133 Roma, Italy\\
$^5$ INFN sezione di Catania and Universit\`a di Catania, 95100 Catania, Italy\\
$^6$ INFN sezione di Genova and Universit\`a di Genova, 16146 Genova, Italy\\
$^7$ IN2P3, Institut de Physique Nucl\'eaire, 91406 Orsay, France\\
$^8$ INFN sezione di Torino  and Universit\`a di Torino, 10125 Torino, Italy\\
$^9$ INFN sezione di Roma I and Istituto Superiore di Sanit\`a, 00161 Roma, Italy\\
$^{10}$ Institute for Nuclear Research, 117312 Moscow, Russia\\
$^{11}$ INFN Laboratori Nazionali di Frascati, 00044 Frascati, Italy\\
$^{12}$ RRC ``Kurchatov Institute", 123182 Moscow, Russia\\
%}

\vspace{0.2in}

{\bf Abstract} - We have studied the angular fluctuations in the speed  of light with
respect to the apex of the dipole of Cosmic Microwave Background (CMB) radiation using
the experimental data obtained with GRAAL facility, located at the European Synchrotron
Radiation Facility (ESRF) in Grenoble.  The measurements were based on the  stability of
the Compton edge of laser photons scattered on the 6~GeV monochromatic electron beam.
The results enable to obtain a conservative constraint on  the anisotropy in the light
speed  variations $\Delta{c(\theta)/c}<3\times 10^{-12}$, i.e. with higher precision than from
previous experiments.

\section{Introduction}

The study of the light speed anisotropy with respect to the dipole of the Cosmic
Microwave Background  (CMB) radiation as suggested in \cite{GM}, is the modern analog of
the Michelson-Morley experiment.  CMB, besides being a unique cosmological messenger,
also is determining the hierarchy of inertial frames  and their relative motions, and is
defining an "absolute" inertial frame of rest, i.e. the one where the  dipole and
quadrupole anisotropies vanish.

The dipole anisotropy of the temperature $T$ of CMB without doubts is of Doppler nature,
\begin{eqnarray}
\frac{\delta T(\theta)}{T}=(v/c)\, cos \theta + (v^2/2c^2)\, cos2\theta + O(v^3/c^3)
\end{eqnarray}
(the first term in the right hand side is the dipole term) and is indicating the
Earth's motion with velocity
$$v/c = 0.000122 \pm  0.00006; v = 365 \pm 18\, km\, s^{-1},$$
with respect to the above mentioned CMB frame.

WMAP satellite's 1-year data define the amplitude of
the dipole $3.346 \pm 0.017$ mK and the coordinate of the apex of the motion  (in
galactic coordinates) \cite{Ben03}
\begin{equation}
l=263.85^{\circ} \pm 0.1^{\circ},
b=48.25^{\circ}\pm 0.04^{\circ}.
\end{equation}
The errors are due to the calibration of
the WMAP and will be improved in due course. This coordinate is in agreement with
estimations based on the hierarchy of motions involving the Galaxy,  Local group and
Virgo supercluster \cite{RG}.

The probing of {\it the anisotropy of the speed of light with
respect to the direction of CMB dipole,} therefore, is a profound
aim.

An experiment was proposed in \cite{GM} to check such a light speed anisotropy using the
effect of  inverse Compton scattering of photons on monochromatic electron beams.
Originally the idea  of using  the Compton scattering of laser photons on accelerated
electrons has been proposed in \cite{HTM},.  Estimations showed the feasibility for
reaching high accuracies at available parameters of laser beams, as well as for highly
monochromatic electrons produced in existing accelerators, such as SLAC, ESRF (Grenoble),
TJNAF. Besides of methodical difference with other experiments such a test would reach
accuracies higher than the existing limits.

Majority of performed measurements of the light speed isotropy were dealing with   a
closed path propagation of light (see \cite{W} and references therein). Such round-trip
propagation are insensitive to the first order but are sensitive only to the  second
order of the velocity of the reference frame of the device with  respect to a
hypothetical universal rest frame.  Mossbauer-rotor experiments yield a one-way limit
$\Delta{c/c}<2\times 10^{-10}$, using fast beam  laser spectroscopy \cite{R}.  The latter
using the light emitted by the atomic beam  yield a limit $\Delta{c/c}<3\times10^{-9}$
for the anisotropy of the one-way  velocity of the light. Similar limit was obtained by
Vessot et al. \cite{V} for the difference in speeds of the uplink and the downlink
signals used in the NASA GP-A rocket experiment to test the gravitational redshift
effect. One-way measurement of the velocity  of light  has  been  performed using also
NASA's Deep Space Network  \cite{KRISH}: the obtained limits  yield
$\Delta c/c <3.5\times 10^{-7}$  and $\Delta c/c <2\times10^{-8}$ for  linear and
quadratic dependencies, respectively. Another class of experiments dealt not with angular
but frequency dependence of the speed of  light. For the visible  light and $\gamma$-rays
the limit $(1.8\pm6)\times 10^{-6}$ was obtained, while the difference with the velocity
of 11~GeV energy electrons $v_e$ was  $(c-v_e)/c=(-1.3\pm2.7)\times10^{-6}$ (see
\cite{MAC}). The data on gamma ray bursters, sources  located on cosmological distances,
were used by Schaefer \cite{Brad} to obtain strong limits  for the variation of the speed
of light on frequency, namely $\Delta c (\nu) /c < 6.3 \times 10^{-21}$ was obtained  for
30-200 keV photons of the flare GRB 930229.

Below we present the first results of the test of the isotropy of the  one-way
speed of light by means of the Compton effect \cite{GM} using the  data obtained at
GRAAL. The results obtained are not only methodically different from those of the above
mentioned experiments but also provide stronger constraints  on the light speed
anisotropy in CMB frame.

\section{Compton Scattering of Photons on Monochromatic Electrons}

In the case of head-on Compton scattering of low energy photons and ultra-relativistic
electrons, the energy dependence of the scattered photon versus angle in the
laboratory frame is given as
\begin{eqnarray}
E_{\gamma}=\frac{4\gamma ^{2}E_{{\ell}}} {1+\frac{\displaystyle 4\gamma
E_{\ell}}{\displaystyle m_{e}}+\theta ^{2}
\gamma ^{2}},
\end{eqnarray}
\noindent where $\theta$ is the angle between the scattered photon and the
incident direction of the electron, $\gamma$ is the Lorentz factor of the
electron (11820 for 6.04~GeV electrons), $E_{\ell}$ is the energy of the incident
photon and m$_{e}$ is the mass of the electron. For visible light (514.5~nm)
the maximum energy of the scattered photons ($\theta =0$, usually called Compton Edge, CE) is at 1100~MeV,
while for the u.v. light (several lines around 351~nm) the CE is at 1500~MeV.

The Compton scattered electrons of energy
$E'_{e}=E_{e}-E_{\gamma}$ will separate from the main beam while
moving through the next magnetic dipole (see the experimental
set-up for further details). The energy of the gamma photons is
related to the distance $x$ between both trajectories by an
equation
\begin{eqnarray}
E_{\gamma}=\frac{E_{e} x}{A+x},
\end{eqnarray}

where A is a constant. For the CE, the equations (3) and (4) lead
to
\begin{eqnarray}
x_{CE}=\frac{4A\gamma E_{\ell}}{m_{e}}.
\end{eqnarray}

The  distance $x_{CE}$ between CE scattered electrons and the main beam is then
proportional to the energy $E_{e}$ of the electrons in the machine.

If the energy of the ultrarelativistic monochromatic electron beams is kept stable
with given accuracy, namely the beam energy is stable over a long time and
not at an instantaneous measurement, the Compton edge variation will result in
the estimation of corresponding light speed variation
\begin{eqnarray}
\beta d\beta = (1/\gamma^2) d\gamma/\gamma.
\end{eqnarray}
\noindent E.g. for $E= 6.04\, GeV$ and $\sigma_E/E \approx 10^{-4}$  one obtains
$$d\beta \leq 10^{-12}.$$

\section{Experimental set-up}

In the experiment carried-out with the GRAAL facility, installed
at the European Synchrotron Radiation Facility (ESRF), Grenoble,
the $\gamma$-ray beam was produced by Compton scattering of laser
photons off the 6~GeV electrons circulating in the  storage ring
($\approx$ 1~km circumference), over a 6.5~m long straight section
\cite{JPB},\cite{NED}. The angle between the incident laser light
and the electron beam does not exceed 1$^{\circ}$, given the
geometry of the set-up, and has no measurable influence on the
beam energy.

\begin{figure}
\begin{center}
\includegraphics[width=0.5\linewidth]{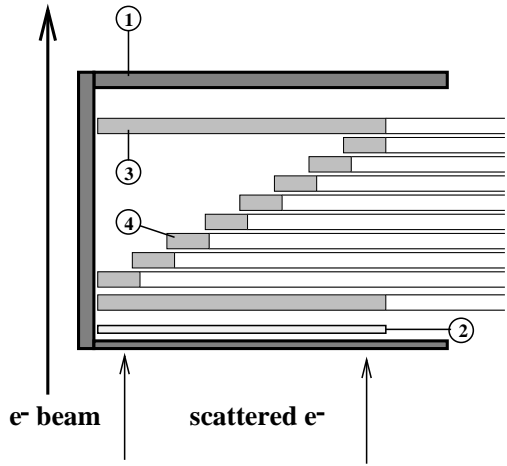}
\caption{Schematic set-up of the tagging box: 1) X-ray shielding, 2) Si microstrip
detector, 3)-4) Scintillators and light guides for timing.}
\end{center}
\end{figure}

\noindent Photon energy was provided by an internal tagging system
located right after the exit dipole of a straight section (Figure
1). Electron position was measured by a silicon microstrip
detector (128 strips with a pitch of 300~$\mu$m  and 500~$\mu$m
thick). Two long plastic scintillators covering the whole focal
plane (38.4~mm) and 8 small ones, each covering  $\approx$1/8~th,
were situated behind the microstrip detector and provided a fast
signal capable of separating 2 electrons of adjacent bunches
(2.8~ns difference) and used as START signal in all Time of Flight
(ToF) measurements of the experiment. The whole system was
embedded in a 4$\pi$ shielding box (equivalent to 8~mm Pb) against
the huge X-ray back-ground. The thickness of this box and the
minimum distance allowed to the electron beam determined our
threshold of 650~MeV in gamma-ray energy.

\begin{figure}
\begin{center}
\includegraphics[width=0.8\linewidth]{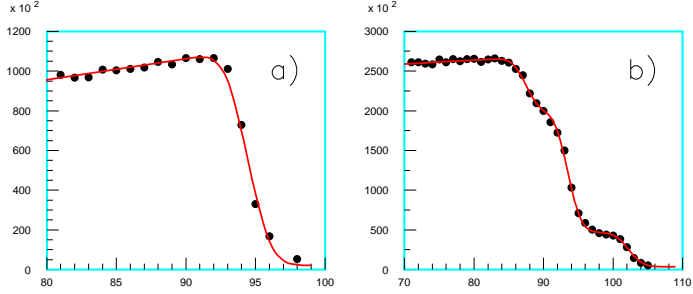}
\caption{Analysis of the Compton edge for visible laser light (a) and u.v.light (b).}
\end{center}
\end{figure}

The CE measured by the silicon microstrip detector is displayed on Fig.2 both
for the visible and u.v. light provided by the laser. The CE  is fitted for a
given laser wave length  by the function
\begin{eqnarray}
N(X)=a_1(1+a_5X)\left(1-{\rm erf}(X)\right)+a_4
\end{eqnarray}
for $X<0$, and
\begin{eqnarray}
N(X)=a_1\left(1-{\rm erf}(X)\right)+a_4
\end{eqnarray}
for $X>0$, where $X=\frac{x-a_2}{a_3 \sqrt{2}}$
and where $x$ is the position (expressed in  microstrip number).
The parameters $a_1$ to
$a_5$ are fitted to the experimental spectrum, $a_2$ represents the CE position
and $a_3$ the Gaussian parameter  $\sigma$, $a_1$ is the amplitude, $a_3$ is the slope of
the spectrum below the  (CE) and $a_4$ is the residual gamma spectrum after the
CE. The residual gamma spectrum is due to the Bremstrahlung of the  electron
beam.
For u.v. three different groups of laser lines: 334.4, 351.1 and 335.8 nm, are present and
can be seen on the experimental  spectrum.

As shown in Figure 2, the experimental energy resolution can be
extracted from the fit of the Compton edge, the theoretical
distribution ending abruptly. In the case of u.v. lines (Figure
2b) the three edges  are clearly resolved and their peak position,
width and relative intensity can be obtained. The measured
resolution (FWHM$\simeq$16~MeV) is in agreement with the estimated
one from the simulation and is dominated by the electron beam
energy dispersion (FWHM=14~MeV). The peak position is used to
calibrate and monitor the absolute position of the microstrip
detector whereas the relative laser line intensities are input to
the calculation of the $\gamma$-ray beam polarization.

Typically an accuracy of $\simeq 0.03 $ microstrip is
obtained for the position of a given line (parameter $a_2$ ). This value  corresponds to
$\simeq 0.2 MeV$ and thus
\begin{eqnarray}
\Delta{ E_{\gamma}}/ E_{\gamma} \simeq 2 \times  10^{-4}.
\end{eqnarray}
As a result 2075 measurements have been performed from 10.04.1998
to 11.05.2002 split in 48 periods of  measurements.
%\end{normalsize}

\section{Data Analysis}

A special interactive software has been developed for the analysis of the GRAAL data. The
data in  our disposal have been represented via sequence of blocks with equal values of
the Compton edge,  as shown in Table 1.

\tabcolsep=0.16em
\begin{table}[h]
    \footnotesize

        \begin{tabular}{ccccccccccccccccc}
             \hline \\
              Block&CE&Laser&Date of&Total&\multicolumn{12}{c}{Months(1998-2002)/Number of
              measurements}\\[1pt]
          &position&nm&measurements&points&I&II&III&IV&V&VI&VII&VIII&IX&X&XI&XII\\[1pt]
             \hline
             \hline    \\
             1& 54.7-56.5 &514.5 &05.06.1999-&389&26& 39& -& - & - &151& -& -& 94&79& - &-\\
              &           &      &05.02.2002 &   &  &   &  &   &   &   &  &  &   &  &   & \\[1pt]
             2& 94.2-94.8 &351.1 &10.04.1998-&443& -& - & -& 87& 62& 64&32&60&138& -& - &-\\
              &           &      &21.09.1998 &   &  &   &  &   &   &   &  &  &   &  &   & \\[1pt]
             3&101.3-101.9&351.1 &16.04.1999-&316& -& - & -&192&124& - & -& -& - & -& - &-\\
              &           &      &16.05.1999 &   &  &   &  &   &   &   &  &  &   &  &   & \\[1pt]
             4&104.4-104.9&351.1 &30.01.2000-&209& 6&145&58& - & - & - & -& -& - & -& - &-\\
              &           &      &06.03.2000 &   &  &   &  &   &   &   &  &  &   &  &   & \\[1pt]
             5&108.0-110.7&334.4,&15.04.2000-&329& -& 39&69& 97&  8& - & -& -& - & -&116&-\\
              &           &351.1 &12.03.2002 &   &  &   &  &   &   &   &  &  &   &  &   & \\[1pt]
             \hline\\
             Total&53.1-110.7&&10.04.1998-&2075&32&261&180&546&256&215&32&60&298&79&116&-\\
                  &          &&11.05.2002 &    &  &   &   &   &   &   &  &  &   &  &   & \\[1pt]
           \hline
      \end{tabular}
      \caption{The dates of the different measurements which contribute to the
      final analysis.}
\end{table}
\normalfont For each block the extreme values, the dates of the
start and the end of measurement are given in Table 1.  The table
shows also the distribution of the periods of measurements over
months.  All existing data have been studied by means of the
normalization by dividing the Compton edge values of each of 48
initial data fragments. Variations both within 24 hour day and
sidereal day have been studied.

\begin{figure}
\begin{center}
\includegraphics[width=0.8\linewidth]{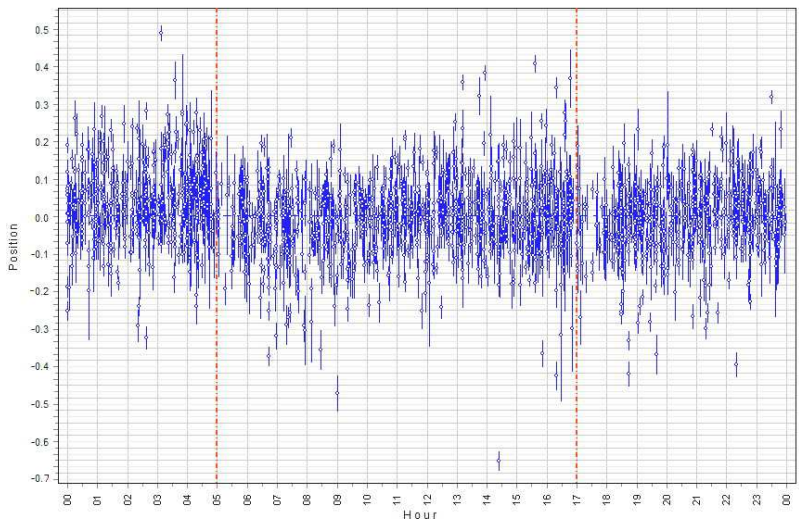}
\caption{
 Experimental data plotted as a function of hour, showing their daily variation.
The dispersion of data around the average, taken arbitrarily at zero, is expressed in
fractions of microstrip (300 micrometers width or about 7 MeV for one microstrip).
  }
\end{center}
\end{figure}

\bigskip

Though large dispersion is apparent, see Figure 3, most of the  experimental points are
located within $\pm 0.3$ microstrip and a stability with accuracy
$2\times 10^{-3}$ can be estimated from the rough data.

\begin{figure}
\begin{center}
\includegraphics[width=0.8\linewidth]{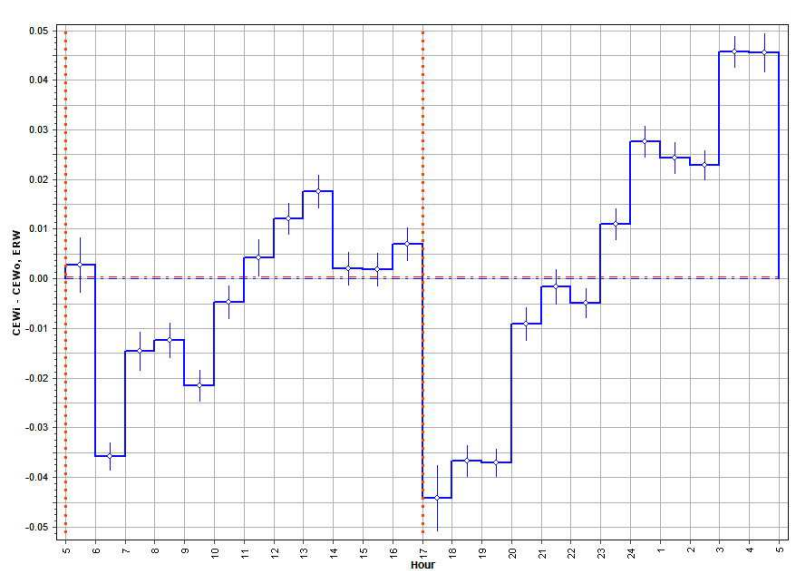}
\caption{
 The same as Figure 3, but each point is the average over one hour. The dotted lines show
the refill time of the machine corresponding to a possible change in temperature of the tagging detector.
The average is also expressed in microstrip fractions.
}
\end{center}
\end{figure}

\bigskip

Figure 4 shows the daily variations of data averaged over 1 hour
intervals. The quasi-monotonous increase of the average energy
after the filling every 12 hour (vertical bars) is clearly
distinguished.  This effect could be understood as a change in
temperature of the  beam vessel and therefore a change in position
of the tagging box with respect to the beam. The correlation is
obvious, but at a low level: 0.09 microstrip over 12 hours i.e.
three times the  uncertainty on a single measurement. However,
this effect is averaged out at our main goal, i.e. when we look at
the data as a function of CMB dipole position.

\begin{figure}
\begin{center}
\includegraphics[width=0.8\linewidth]{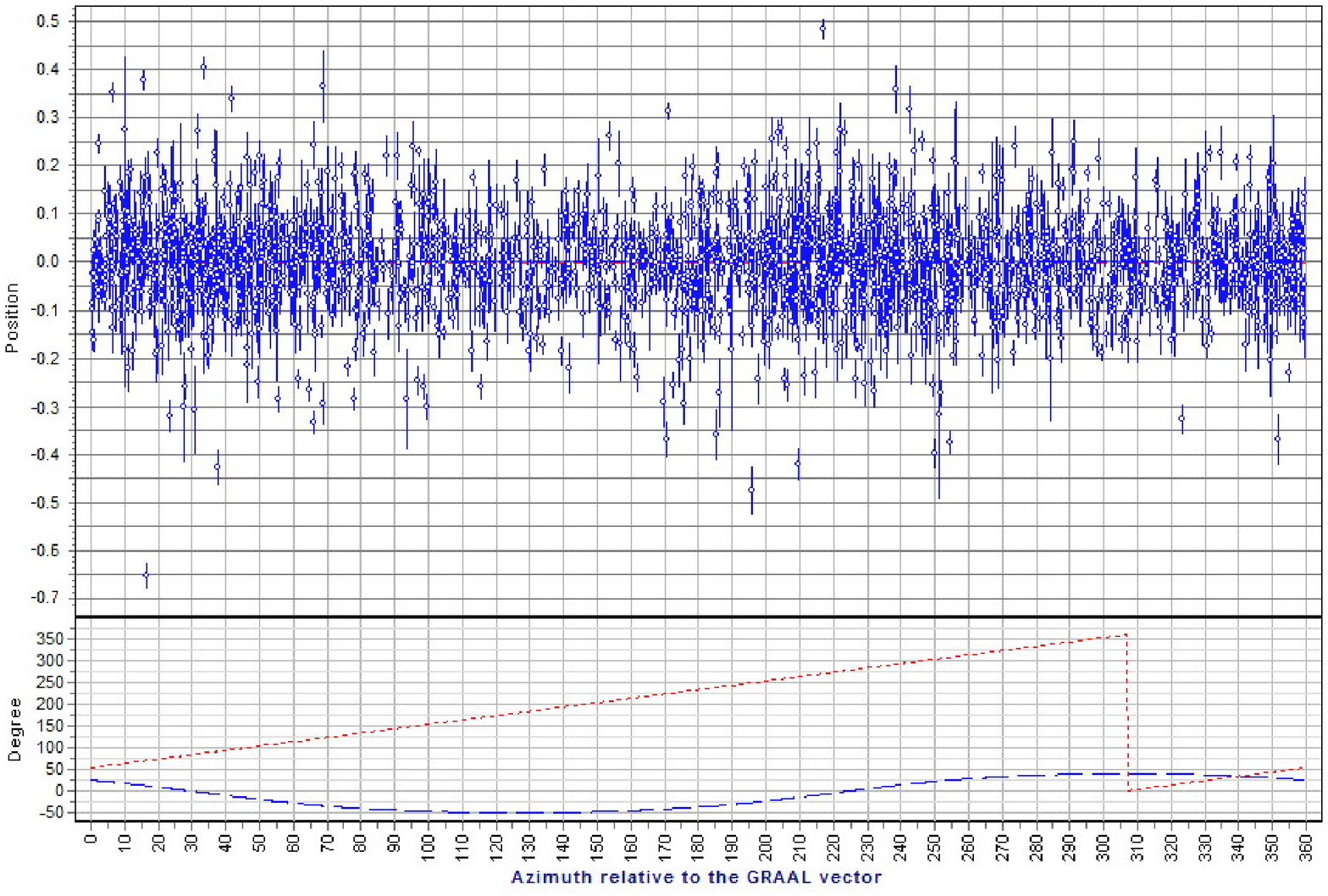}
\caption{ Experimental data plotted as a function of the azimuth
(above); below, the variation of the angle between the beam and
the CMB dipole decomposed to azimuth (dotted) and declination
(dashed) angles is shown. Dispersion has the same expression as
for the two previous figures.}
\end{center}
\end{figure}

\bigskip

The variation of the data over the azimuth angle is shown in
Figure 5, together with the variation of the angular distance
between the beam and the apex of the CMB dipole. Note the
existence of up to 10 $\sigma$ fluctuations for the individual
points which we could not correlate to any systematic effect.

\begin{figure}
\begin{center}
\includegraphics[width=0.8\linewidth]{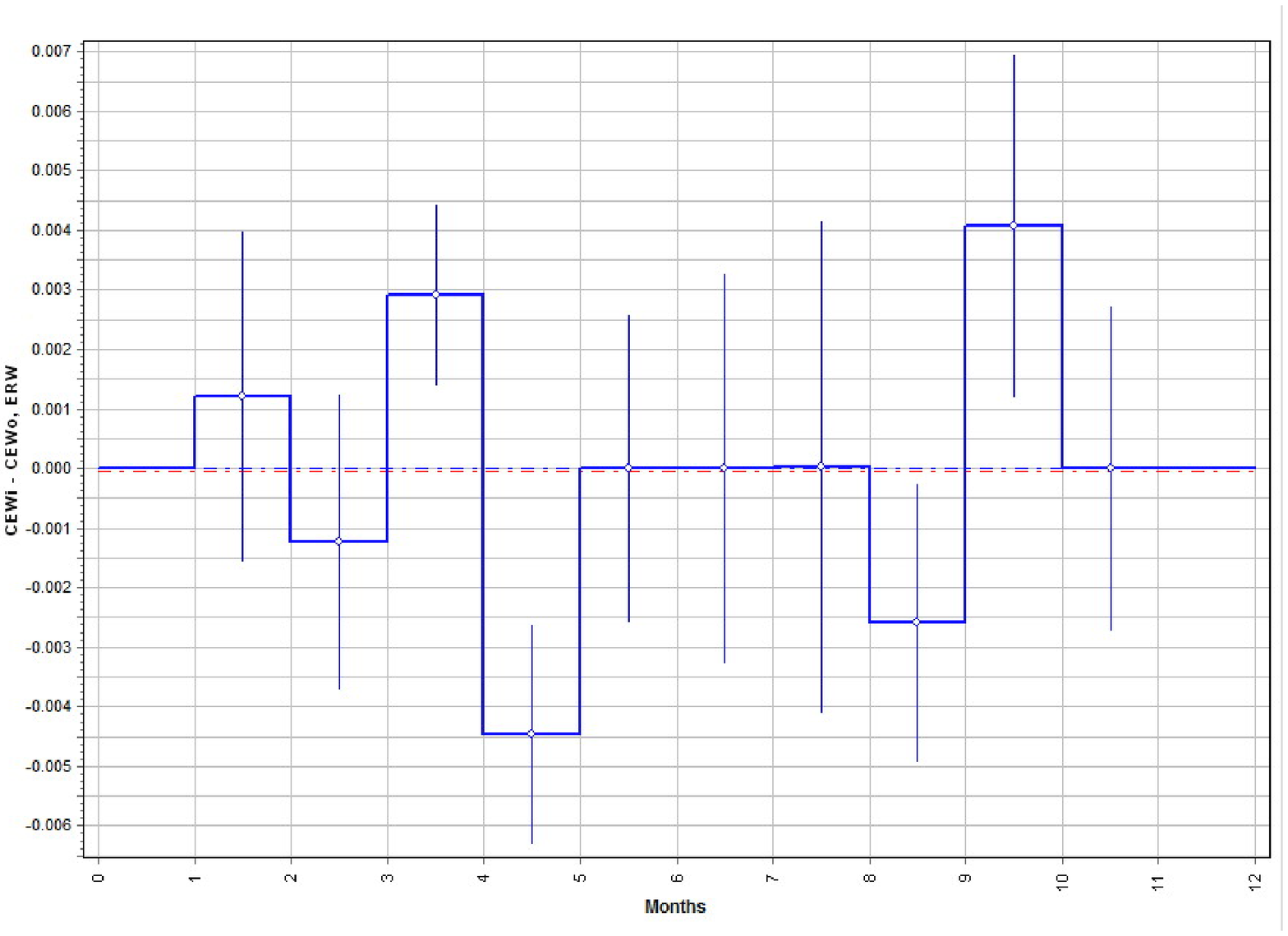}
\caption{
 Experimental data over 12 month period averaged by month. The average, also expressed in
microstrip fractions, is found very close to zero.
}
\end{center}
\end{figure}

 The distribution of all data averaged by month is given in
Figure 6. As can be seen in that plot, the averaging procedure
gives very low values for the fluctuations, showing that the daily
fluctuations observed in Figure 4 are averaged out. From this plot
we can estimate the fluctuations to $\simeq  0.015 \pm 0.007$
microstrip corresponding to
\begin{eqnarray}
\Delta E_{\gamma} /E_{\gamma} \simeq (1 \pm 0.5) \times 10^{-4}.
\end{eqnarray}
 The  uncertainty  is statistical only and reflects the averaging procedure.

\section{Conclusions}

From the stability limit shown in Figure 6, we established the
following constraint for the light speed anisotropy: $ \Delta c/c
\leq (1 \pm 0.5) \times  10^{-12}. $ The systematic errors of the
experiment are indeed very difficult to estimate, thus in view of
the large and unexplained fluctuations of the fitted value for the
CE, we adopt a conservative limit
\begin{eqnarray}
\Delta c/c < 3 \times  10^{-12}.
\end{eqnarray}

However, even this conservative limit is about 3 orders of
magnitude better than the one reachable from measurements
involving the space probes (including the Cassini probe), since it
will correspond to several meter accuracy for a planetary orbit;
also it deals with one-way effect, while the space measurements
are two-way ones (Nordtvedt, personal communication, 2004).

It is important to note that, the electrons are circulating inside the machine at  a rate
of $10^6$ revolutions per second,  and therefore, an anisotropy effect cannot be compensated
by the stabilization procedure of the machine as it could happen in a linear accelerator.
In  other words, the angle between the electron speed  inside the ring and the direction
of the CMB dipole is changing so fast, that the machine cannot compensate for a change in energy
induced by an anisotropy.

Further studies of the light speed anisotropy with respect to the
apex of CMB dipole, possibly in dedicated experiments with higher
$\gamma$ and lower $d\gamma$, seem crucial and can reduce our
limit or give evidence for possible anisotropy effects.

We thank K.Nordtvedt for valuable discussions and the ESRF teams for their
permanent help, in particular, L.Hardy, J.L.Revol, D.Martin and the members of the alignment group.

\end{normalsize}


\begin{thebibliography}{99}

\bibitem{GM}
Gurzadyan V.G., Margarian A.T., Physica Scripta, {\bf 53}, 513, 1996.

\bibitem{Ben03} Bennett C.L., et al., ApJ Suppl. {\bf 148}, 1, 2003.

\bibitem{RG}
Rauzy S., Gurzadyan V.G., Mon. Not. Roy. Astr. Soc., {\bf 298}, 114, 1998.

\bibitem{HTM}
Harutyunian F.R., Tumanian V.A. Phys.Lett. {\bf 4}, 176, 1963; Milburn R.H., Phys.Rev.Lett. {\bf 10}, 75, 1963.

\bibitem{JPB}
Bocquet J.P. et al. Nucl. Phys. {\bf A622}, 125, 1997.

\bibitem{NED}
Nedorezov V.G., Turinge A.A., Shatunov  Yu.M. Physics-Uspekhi, {\bf 47}, 341, 2004.

\bibitem{W}
Will C.M. Phys.Rev.D., {\bf 45}, 403, 1992.

\bibitem{R}
Riis E. et.al. Phys.Rev.Lett. {\bf 60}, 81, 1988;
Bay Z. and White J., ibid. {\bf 62}, 841, 1989.

\bibitem{V}
Vessot et al.,Phys.Rev.Lett. {\bf 45}, 2081, 1980.

\bibitem{KRISH}
Krisher T.P et al.,Phys.Rev.D {\bf 42}, 731, 1990.

\bibitem{Brad}
Schaefer B.E. Phys. Rev. Lett. {\bf 82}, 4964, 1999.

\bibitem{MAC}
MacArthur D.W. Phys.Rev.A {\bf 33}, 1, 1986.


\end{thebibliography}
\end{document}